# Multipath Time-delay Estimation with Impulsive Noise via Bayesian Compressive Sensing

Xingyu Ji, Lei Cheng, and Hangfang Zhao

*Abstract*—Multipath time-delay estimation is commonly encountered in radar and sonar signal processing. In some real-life environments, impulse noise is ubiquitous and significantly degrades estimation performance. Here, we propose a Bayesian approach to tailor the Bayesian Compressive Sensing (BCS) to mitigate impulsive noises. In particular, a heavy-tail Laplacian distribution is used as a statistical model for impulse noise, while Laplacian prior is used for sparse multipath modeling. The Bayesian learning problem contains hyperparameters learning and parameter estimation, solved under the BCS inference framework. The performance of our proposed method is compared with benchmark methods, including compressive sensing (CS), BCS, and Laplacian-prior BCS (L-BCS). The simulation results show that our proposed method can estimate the multipath parameters more accurately and have a lower root mean squared estimation error (RMSE) in intensely impulsive noise.

## I. Introduction

MULTIPATH time-delay estimation, also known as time of arrival (TOA), is a fundamental concept with diverse applications in radar [1], sonar [2], and wireless communication systems [3]. A significant amount of literature has been dedicated to the topic of TOA estimation.

Classical methods for TOA estimation include matched filter (MF) [4] and subspace methods such as multiple signal classification (MUSIC) algorithm [5]. Most recently, compressive sensing (CS) has emerged as a promising for achieving higher resolution [6]. The convex optimization method for *l*-1 norm minimization using the CVX toolbox is widely used [7]. Another powerful solver for CS is sparse Bayesian learning (SBL) [8]–[10], also known as Bayesian compressive sensing (BCS) [11]. BCS-based methods utilize the flexible probability model framework, showing highly effective in CS-TOA estimation [1][2]. While BCS uses Gaussian prior, L-BCS [12] uses Laplacian prior in order to further promote sparsity.

Those aforementioned BCS methods are derived based on Gaussian noise. However, in real environments, impulse noise is often encountered in compressed sensing applications such as sparse channel estimation in powerline communications [13], underwater communications [14], ships detection [15], and ocean acoustic tomography [16]. While the use of BCS for impulse noise modeling was studied in [17], there has been a lack of research on the combination of CS-TOA estimation with impulsive noise, particularly from the perspective of BCS.

This letter proposes a probabilistic model that uses Laplacian likelihood and Laplacian prior. It is an extension of BCS with Gaussian likelihood and Gaussian prior [9][12]. The Laplacian likelihood is to model impulse noise using the heavy-tail property of Laplacian distribution. The purpose of the Laplacian prior is to promote the sparsity of signal [12][18]. The posterior distribution of time delay is derived where the involved hyper-parameters are updated under the expectation maximization (EM) framework and Type-II likelihood approximation [12]. We refer to this new method as LL-BCS due to its use of Laplacian likelihood and Laplacian prior in the BCS framework. LL-BCS was found to outperform most state-of-the-art algorithms in the presence of intensely impulsive noise, as demonstrated by simulation results.

The rest of this letter is organized as follows. Section II presents the CS model for solving the multipath time-delay estimation problem. The reconstruction algorithm is derived in Section III. Section IV presents simulation results. Finally, conclusions are drawn in Section V.

## II. Signal Model and Problem Formulation

Assuming $K$ multipaths in a sonar or radar channel, the impulse response function $h(t)$ can be modeled as

$$h(t) = \sum_{k=0}^{K-1} \alpha_k \delta(t - \tau_k), \tag{1}$$

where $\alpha_k$ and $\tau_k$ are the complex amplitude and time delay of

This work was supported in part by the National Natural Science Foundation of China under Grant Nos. 62071429 and 62001309, in part by the National Key R&D Program of China under Grant 2016YFC1400100, in part by Science and Technology on Sonar Laboratory under Grant 6142109KF212204, and in part by the Fundamental Research Funds for the Central Universities and Zhejiang University Education Foundation Qizhen Scholar Foundation. (Corresponding author: Hangfang Zhao)

Xingyu Ji, Lei Cheng, and Hangfang Zhao are with the College of Information Science and Electronic Engineering, Zhejiang University, Hangzhou City, Zhejiang Province 310027, China (e-mail: ji_xing_yu@zju.edu.cn; lei_cheng@zju.edu.cn; hfzhao@zju.edu.cn).

Lei Cheng is also with Zhejiang Provincial Key Laboratory of Information Processing, Communication and Networking (IPCAN), Hangzhou 310027, China.

Hangfang Zhao is also with Laboratory of Ocean Observation-Imaging Testbed of Zhejiang Province, Zhoushan City, Zhejiang Province 316021, China.

the $k$-th path, respectively. $\delta(t)$ is the Dirac delta function. Through the multipath channel, the received echo signal $y(t)$ can be expressed as

$$y(t) = s(t) * h(t) + e(t)$$
$$= \sum_{k=0}^{K-1} \alpha_k s(t - \tau_k) + e(t) \quad 0 \le t \le T_s + \tau_{K-1}, \quad (2)$$

where $s(t)$ is the broadband transmitting/radiated signal, $e(t)$ represents additive noise, $T_s$ denotes signal duration, and $\tau_{K-1}$ is the latest TOA.

We will now proceed to represent the above equations in discrete form. With sampling frequency $f_s$, $s(t)$ can be denoted as the discrete transmitting signal $s[n]$, $n \in \{0, ..., I-1\}$, where $T_s \times f_s + 1 = I$ is the transmitting signal length. The discrete impulse response function is $h[n] = \sum_{k=0}^{K-1} \alpha_k \delta[n - T_k]$, where discrete time delay points $0 < T_0 < T_1 < ... < T_{K-1}$ correspond to $0 < \tau_0 < \tau_1 < ... < \tau_{K-1}$. The received signal can be expressed in discrete form, given by

$$y[n] = s[n] * h[n] + e[n]$$
$$= \sum_{k=0}^{K-1} \alpha_k s[n - T_k] + e[n] \quad 0 \le n \le I - 1 + T_{K-1}, \quad (3)$$

The next step is to define the TOA search length and resolution for the above equation. Define $\tau_{max}$ as the maximal possible time delay bound, and naturally, it follows that $\tau_{K-1} < \tau_{max}$. The search resolution is defined as the sampling rate $f_G$, which is a down-sampling from $f_s$. Generally, $f_s / f_G$ is set as an integer. If $f_s / f_G$ is not an integer, the sinc convolution expression can be found in [19]. Thus, the search length is defined as $\tau_{max} \cdot f_G = N$. Eq.(3) can be converted to

$$y[n] = \sum_{i=0}^{N-1} h[i] s[n-i] + e[n] \quad 0 \le n \le M, \quad (4)$$

where we define $M \triangleq (T_s + \tau_{max}) \cdot f_s$, and use the notational convention that $s[n] = 0$ for $n \notin \{0, 1, 2, ..., I\}$.

Then, the above expressions can be shown in matrix form,

$$\mathbf{y} = \mathbf{A}\mathbf{x} + \mathbf{e}, \quad (5)$$

where the vectors of the impulse response function $\mathbf{x}$, receiver signal $\mathbf{y}$, and noise $\mathbf{e}$, expressed as

$$\mathbf{x} = [h[0], h[1], ..., h[N]] \in \mathbb{R}^{N \times 1}, \quad (6)$$
$$\mathbf{y} = [y[0], y[1], ..., y[M]] \in \mathbb{R}^{M \times 1}, \quad (7)$$
$$\mathbf{e} = [e[0], e[1], ..., e[M]] \in \mathbb{R}^{M \times 1}, \quad (8)$$

respectively and the sensing matrix $\mathbf{A}$ is

$$\mathbf{A} = \begin{bmatrix} s[0] & & 0 \\ ... & ... & ... \\ s[I] & ... & s[0] \\ ... & ... & ... \\ 0 & & s[I] \end{bmatrix} \in \mathbb{R}^{M \times N}. \quad (9)$$

There are $K$ non-zero values corresponding to $K$ multipath in $\mathbf{x}$ and others zeros. Naturally, the search length grids are much larger than the number of multipath, i.e. $N \gg K$, which reflects the sparsity of $\mathbf{x}$.

### III. PROPOSED METHOD

Building on the above derivation, the sparsity of the multipath time-delay estimation model is explicitly defined. We will introduce Laplacian distribution for both the prior and likelihood to model Eq.(5) in a probabilistic way. Subsequently, a BCS-based algorithm will be derived to solve the probabilistic inference problem.

#### A. Likelihood and Prior

For the noise $\mathbf{e}$, we assume it is independent and identically distributed (i.i.d.) Laplacian noise with zero mean and variance equal to $2\beta^{-1}$. The data likelihood for channel impulse response $\mathbf{x}$ and observation time-domain signal $\mathbf{y}$ is Laplacian as

$$p(\mathbf{y} | \mathbf{x}, \beta) = \mathcal{L}(\mathbf{y} | \mathbf{A}\mathbf{x}, \beta^{-1}\mathbf{I}) = \prod_{i=1}^{M} \mathcal{L}(y_i | [\mathbf{A}\mathbf{x}]_i, \beta^{-1})$$
$$= \prod_{i=1}^{M} \frac{\sqrt{\beta}}{2} \exp\left(-\sqrt{\beta} \|y_i - [\mathbf{A}\mathbf{x}]_i\|_1\right). \quad (10)$$

The key observation for our further algorithm design is that we can represent the Laplacian distribution as an integral of Gaussian mixture models (GMM) [20] as follows

$$\mathcal{L}(y_i | [\mathbf{A}\mathbf{x}]_i, \beta^{-1}) = \int p(y_i | \tau_i) p(\tau_i | \beta) d\tau_i$$
$$= \int \mathcal{N}(y_i | [\mathbf{A}\mathbf{x}]_i, \tau_i) \Gamma\left(\tau_i | 1, \frac{\beta}{2}\right) d\tau_i, \quad (11)$$

where $\Gamma(\zeta | a, b) = b^a / \Gamma(a) \zeta^{a-1} \exp[-b\zeta]$. Then substitute Eq.(11) into Eq.(10), it will be

$$p(\mathbf{y} | \mathbf{x}, \beta) = \prod_{i=1}^{M} \int \mathcal{N}(y_i | [\mathbf{A}\mathbf{x}]_i, \tau_i) \Gamma\left(\tau_i | 1, \frac{\beta}{2}\right) d\tau_i$$
$$= \mathcal{N}(\mathbf{A}\mathbf{x}, \Sigma^y) \prod_{i=1}^{M} \Gamma\left(\tau_i | 1, \frac{\beta}{2}\right). \quad (12)$$

where the received signal $\mathbf{y}$ follows multivariate Gaussian distribution $\mathcal{N}(\mathbf{A}\mathbf{x}, \Sigma^y)$ in the core hierarchical likelihood, $\boldsymbol{\tau} = [\tau_1, \tau_2, ..., \tau_M]^T$ is the noise variance, and $\Sigma^y = \text{diag}(\boldsymbol{\tau})$ is the covariance matrix. The proposed model constitutes a two-stage hierarchical form. The hierarchical priors in Eq.(12) decompose the Laplacian distribution into weighted Gaussian priors, which will be helpful in posterior calculation and Bayesian inference.

For the prior, we assume that the multipath amplitudes $\alpha_k$ are independent of each other and follow a Laplacian distribution with different variances. Here the Laplacian distribution is mainly for promoting the sparsity of $\mathbf{x}$ [12][18]. Then

$$p(\mathbf{x} | \lambda) = \mathcal{L}(\mathbf{x} | \mathbf{0}, \lambda^{-1}\mathbf{I}) = \prod_{j=1}^{N} \mathcal{L}(x_j | 0, \lambda^{-1})$$
$$= \prod_{j=1}^{N} \frac{\sqrt{\lambda}}{2} \exp\left(-\sqrt{\lambda} \|x_j\|_1\right). \quad (13)$$

Using the GMM representation, Eq.(13) can be expressed as

$$p(\mathbf{x}|\lambda) = \int p(\mathbf{x}|\boldsymbol{\gamma})p(\boldsymbol{\gamma}|\lambda)d\boldsymbol{\gamma} = \prod_{j=1}^{N}\int p(x|\gamma_j)p(\gamma_j|\lambda)d\gamma_j$$

$$= \prod_{j=1}^{N}\int \mathcal{N}(x_j|0,\gamma_j)\Gamma\left(\gamma_j|1,\frac{\lambda}{2}\right)d\gamma_j \quad (14)$$

$$= \mathcal{N}(\mathbf{x}|\mathbf{0},\boldsymbol{\Sigma}^x)\prod_{j=1}^{N}\Gamma\left(\gamma_j|1,\frac{\lambda}{2}\right).$$

The impulse response function $\mathbf{x}$ follows multivariate Gaussian distribution $\mathcal{N}(0,\boldsymbol{\Sigma}^x)$ in the core hierarchical priors, where $\boldsymbol{\gamma} = [\gamma_1,\gamma_2,\ldots,\gamma_M]^T$ is channel amplitude variance and $\boldsymbol{\Sigma}^x = \text{diag}(\boldsymbol{\gamma})$ is the covariance matrix.

*B. Posterior*

Due to the core hierarchical distribution terms $\mathcal{N}(\mathbf{y}|\mathbf{Ax},\boldsymbol{\Sigma}^y)$ in Eq.(12) and $\mathcal{N}(\mathbf{x}|\mathbf{0},\boldsymbol{\Sigma}^x)$ in Eq.(14) being Gaussian distribution, $p(\mathbf{x}|\mathbf{y},\boldsymbol{\tau},\boldsymbol{\gamma},\beta,\lambda)$ will be multivariate Gaussian distribution $\mathcal{N}(\mathbf{x}|\boldsymbol{\mu},\boldsymbol{\Sigma})$ with parameters [12]

$$\boldsymbol{\mu} = \boldsymbol{\Sigma}\mathbf{A}^T\left[\boldsymbol{\Sigma}^y\right]^{-1}\mathbf{y}, \quad (15)$$

$$\boldsymbol{\Sigma}^{-1} = \left[\boldsymbol{\Sigma}^x\right]^{-1} + \mathbf{A}^T\left[\boldsymbol{\Sigma}^y\right]^{-1}\mathbf{A}. \quad (16)$$

*C. Hyperparameters Update Solutions*

The posterior distribution $p(\mathbf{x},\boldsymbol{\tau},\boldsymbol{\gamma},\beta,\lambda|\mathbf{y})$ is further decomposed based on the following decomposition

$$p(\mathbf{x},\boldsymbol{\tau},\boldsymbol{\gamma},\beta,\lambda|\mathbf{y}) = p(\mathbf{x}|\mathbf{y},\boldsymbol{\tau},\boldsymbol{\gamma},\beta,\lambda)p(\boldsymbol{\tau},\boldsymbol{\gamma},\beta,\lambda|\mathbf{y}). \quad (17)$$

The latter term $p(\boldsymbol{\tau},\boldsymbol{\gamma},\beta,\lambda|\mathbf{y})$ is replaced by a degenerate distribution in the type-II maximum likelihood procedure as

$$p(\boldsymbol{\tau},\boldsymbol{\gamma},\beta,\lambda|\mathbf{y}) = p(\mathbf{y},\boldsymbol{\tau},\boldsymbol{\gamma},\beta,\lambda)/p(\mathbf{y}) \propto p(\mathbf{y},\boldsymbol{\tau},\boldsymbol{\gamma},\beta,\lambda), \quad (18)$$

where $p(\mathbf{y},\boldsymbol{\tau},\boldsymbol{\gamma},\beta,\lambda)$ is the evidence term, which can be derived from the EM algorithm over $\mathbf{x}$ in $p(\mathbf{y},\boldsymbol{\tau},\mathbf{x},\boldsymbol{\gamma},\beta,\lambda)$. Assuming uniform hyperpriors $p(\beta)$ and $p(\lambda)$ for the parameters $\beta$ and $\lambda$ [21] and combining the hierarchical observation and signal models, the joint distribution takes as

$$p(\mathbf{y},\boldsymbol{\tau},\mathbf{x},\boldsymbol{\gamma},\beta,\lambda) = p(\mathbf{y}|\mathbf{x},\boldsymbol{\tau})p(\mathbf{x}|\boldsymbol{\gamma})p(\boldsymbol{\tau}|\beta)p(\boldsymbol{\gamma}|\lambda)p(\beta)p(\lambda)$$

$$= \mathcal{N}(\mathbf{y}|\mathbf{Ax},\boldsymbol{\Sigma}^y)\prod_{i=1}^{M}\Gamma\left(\tau_i|1,\frac{\beta}{2}\right)\mathcal{N}(\mathbf{x}|\mathbf{0},\boldsymbol{\Sigma}^x)\prod_{j=1}^{N}\Gamma\left(\gamma_j|1,\frac{\lambda}{2}\right). \quad (19)$$

Taking logarithm and expectation of $p(\mathbf{y},\boldsymbol{\tau},\mathbf{x},\boldsymbol{\gamma},\beta,\lambda)$, then we get the expectation of log-evidence $Q$ as

$$Q \equiv \mathbb{E}_{\mathbf{x}}[\ln p(\mathbf{y},\boldsymbol{\tau},\mathbf{x},\boldsymbol{\gamma},\beta,\lambda)] =$$
$$-\frac{1}{2}\sum_{j=1}^{N}\log\gamma_j - \frac{1}{2}\sum_{j=1}^{N}\frac{1}{\gamma_j}(\mu_j^2 + \Sigma_{jj}) - \frac{1}{2}\sum_{i=1}^{M}\log\tau_i$$
$$-\frac{1}{2}\sum_{i=1}^{M}\frac{1}{\tau_i}[y_i^2 - 2y_i(\mathbf{A}\boldsymbol{\mu})_i + [\mathbf{A}(\boldsymbol{\Sigma}+\boldsymbol{\mu}\boldsymbol{\mu}^T)\mathbf{A}^T]_{ii}] \quad (20)$$
$$+M\ln(\frac{\beta}{2}) - \beta\sum_{i=1}^{M}\frac{\tau_i}{2} + N\ln(\frac{\lambda}{2}) - \lambda\sum_{i=1}^{N}\frac{\gamma_i}{2}.$$

Further taking the derivative of $Q$ over $\gamma_j,\lambda,\beta,\tau_i$, we can get the hyperparameters iteratively

TABLE I
LL-BCS ALGORITHM

| |
|---|
| Initialization, $\lambda = 0.1$, $\beta = 0.1$, $\boldsymbol{\gamma} = \mathbf{A}^\dagger\mathbf{y}$, $\epsilon_{\min} = 0.001$, $j_{\max} = 1000$ |
| 1  while ( $\epsilon > \epsilon_{\min}$ ) and ( $j < j_{\max}$ ) |
| 2   $\boldsymbol{\gamma}^{old} = \boldsymbol{\gamma}^{new}$, $\boldsymbol{\tau}^{old} = \boldsymbol{\tau}^{new}$, $\boldsymbol{\Sigma}^x = \text{diag}(\boldsymbol{\gamma}^{new})$, $\boldsymbol{\Sigma}^y = \text{diag}(\boldsymbol{\tau}^{new})$ |
| 3   update $\boldsymbol{\mu}$ and $\boldsymbol{\Sigma}$ using Eq.(15)-(16) |
| 4   update $\boldsymbol{\gamma},\lambda,\beta,\boldsymbol{\tau}$ using Eq.(21)-(24) |
| 5   $\epsilon = \|\boldsymbol{\gamma}^{new} - \boldsymbol{\gamma}^{old}\|_1 / \|\boldsymbol{\gamma}^{old}\|_1$ |
| 6   $j = j+1$ |
| 7 end |
| 8 Output: $\boldsymbol{\gamma}^{new}$ |

$$\gamma_j = \frac{-1}{2\lambda} + \sqrt{\frac{1}{4\lambda^2} + \frac{\mu_j^2 + \Sigma_{jj}}{\lambda}}, \quad j=1,2,\ldots N, \quad (21)$$

$$\lambda = N \bigg/ \sum_{j=1}^{N}\frac{\gamma_j}{2}, \quad (22)$$

$$\beta = M \bigg/ \sum_{i=1}^{M}\frac{\tau_i}{2}, \quad (23)$$

$$\tau_i = \frac{-1}{2\beta} + \sqrt{\frac{1}{4\beta^2} + \frac{R_i}{\beta}}, \quad i=1,2,3\ldots M, \quad (24)$$

where $R_i = y_i^2 - 2y_i[\mathbf{A}\boldsymbol{\mu}]_i + [\mathbf{A}(\boldsymbol{\Sigma}+\boldsymbol{\mu}\boldsymbol{\mu}^T)\mathbf{A}^T]_{ii}$.

*D. LL-BCS Algorithm*

The algorithm of the proposed method is shown in Table I. We iteratively update hyperparameters $\boldsymbol{\gamma},\lambda,\beta,\boldsymbol{\tau}$ using Eq.(21)-(24) and posterior $\boldsymbol{\mu},\boldsymbol{\Sigma}$ using Eq.(15)-(16). The terminating condition of iteration is that the number of iterations reaches the maximum $j_{\max}$ or convergence rate

$$\epsilon = \|\boldsymbol{\gamma}^{new} - \boldsymbol{\gamma}^{old}\|_1 / \|\boldsymbol{\gamma}^{old}\|_1, \quad (25)$$

achieves $\epsilon < \epsilon_{\min}$. The resulting $\boldsymbol{\gamma}$ is a grid-based time-delay spectrum. The sequences of parameters estimation for $\boldsymbol{\gamma},\lambda,\beta,\boldsymbol{\tau}$ in the EM iterations converge in Eq.(21)-(24) [22]. However, convergence is only guaranteed toward a *local* optimum of the marginal log-likelihood $p(\mathbf{y},\boldsymbol{\tau},\boldsymbol{\gamma},\beta,\lambda)$. Guarantees for convergence can be given using Eq.(25), but is hard to prove analytically. In practice, we calculate the initialization with the least square solution $\boldsymbol{\gamma} = \mathbf{A}^\dagger\mathbf{y}$, which is relatively close to the ground truth and improves LL-BCS's stability and convergence speed.

IV. SIMULATIONS

In this section, we perform a time-delay estimation simulation with impulse noise to verify the performance of LL-BCS.

We use the receiver signal processing procedure in Sec.II to establish the measurement equation. The transmitting signal is a linear frequency modulation (LFM) signal with frequencies ranging from 6 kHz to 7 kHz and $T_s = 0.05\ s$, similar in [23]. The signal sampling frequency $f_s$ is 20 kHz. The TOA sampling frequency $f_G$ for the search resolution is set as 2 kHz, which is twice as much as the LFM signal bandwidth. For the following simulations, we set $K = 4$. The $K$ non-zero time-delay multipaths are randomly picked in a time-delay window from 0 to $\tau_{max} = 0.02\ s$, with their amplitudes selected as identity. Thus, $N = \tau_{max} \cdot f_G = 40$, $M = (T_s + \tau_{max}) \cdot f_s = 1400$.

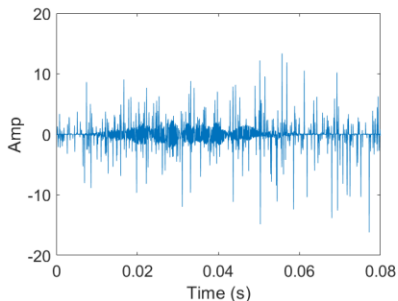

Fig. 1. Received signal waveform (SINR is -10 dB).

We consider the signal reconstruction with impulsive noise. Synthetic impulsive noise was generated from a GMM. The GMM has three Gaussian distributions $\mathcal{N}(0, l_i)$, with corresponding variances $l_i$ and mixing probability $\pi_i$ of the $i$-th Gaussian component. $\pi_i$ acts like $\Gamma(\gamma_i | 1, \lambda/2)$ in Eq.(14). The impulse noise is expressed as

$$e_{impulse}[n] = \sum_{i=1}^{3} \pi_i \cdot \mathcal{N}(0, l_i) \quad 0 \le n \le M. \quad (26)$$

where the parameters are $l_1 = 1$, $l_2 = 10$, and $l_3 = 100$, respectively, and $\pi_1 = 0.9$, $\pi_2 = 0.07$, and $\pi_3 = 0.03$, similar in [13]. The total noise is a sum of impulsive noise and Gaussian noise, as

$$e[n] = e_{Gaussian}[n] + e_{impulse}[n] \quad 0 \le n \le M. \quad (27)$$

First, the signal-to-noise ratio (SNR) is defined in detail. We separately calculate the SNR for Gaussian noise $e_{Gaussian}$ and impulsive noise $e_{impulse}$, denoted as SGNR and SINR, respectively. With SGNR being 20 dB and SINR being -10 dB, Fig.1 shows the received signal waveform in the time domain. In this scenario, the signal is heavily obscured by impulsive noise.

Next, the reconstructed impulse response functions are compared. Our algorithm LL-BCS is evaluated against state-of-the-art sparsity-based algorithms, including the $l$-1 norm convex optimization method by CVX [7], BCS [11], and L-BCS [12]. The SNR values are the same as in Fig.1 for both Gaussian and impulsive noise. Fig.2 shows the results in a scenario with four multipaths. The ground truth values are marked as red dots, and the estimated time-delay results are shown in blue lines, while identified peaks are marked as black circles. It is observed that CVX and L-BCS fail to capture the positions of the multipath, while BCS provides one false time delay peak. LL-BCS achieves an almost perfect estimation of the multipath parameters, both for delay time and amplitude.

Finally, we evaluate the performance of all methods in different SINRs using the Monte Carlo method. We select SINR ranging from -20 dB to 20 dB and SGNR being fixed at 20 dB. RMSE is calculated by

$$RMSE = \frac{1}{\mathcal{M} \cdot K} \sqrt{\sum_{m=0}^{\mathcal{M}-1} \sum_{k=0}^{K-1} (\hat{\tau}_k^{(m)} - \tau_k^{(m)})^2}, \quad (28)$$

where $\mathcal{M}$ denotes the number of independent Monte Carlo simulations. $\hat{\tau}_k^{(m)}$ and $\tau_k^{(m)}$ are the $k$-th estimated time delay and $k$-th time delay ground truth in the $m$-th simulation, respectively. The estimated time-delay $\hat{\tau}_k^{(m)}$ is the $k$-th peak of the maximal $K$ peaks in estimated $\gamma$. From the simulation results using 250 Monte Carlo trials shown in Fig.3, LL-BCS outperforms other methods significantly in low SINR cases. As the SINR increases, the RMSEs of all methods converge to zero, and the gaps between them decrease. In high SINR scenarios, the heavy-tail property of impulsive noise will be insignificant compared to the background Gaussian noise.

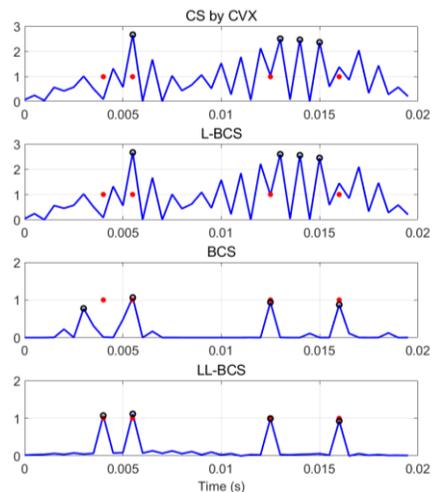

Fig. 2. Time delay estimation results of four methods.

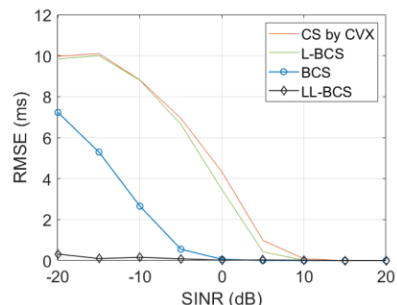

Fig. 3. RMSE results versus SINR.

## V. CONCLUSION

A multipath time-delay estimation method with impulsive noise named LL-BCS is proposed. The multipath time-delay signal model is derived in the time domain. Impulsive noise is modeled as a heavy-tail Laplacian distribution, resulting in a Laplacian likelihood. Relying on its GMM representation, closed-form Bayesian inference steps are derived via Bayesian compressive sensing. Simulation results show that LL-BCS mitigates the outliers introduced by impulsive noise and outperforms other state-of-art competitors under intensively impulsive noise in terms of RMSE for time delay estimation.